\documentclass[apj]{emulateapj}
\usepackage{epsfig}
\newcommand{\Msun}{\ensuremath{\,{\rm M}_\odot}}

\newcommand{\ds}{\displaystyle}
\newcommand{\kms}{km\,s$^{-1}$}
\newcommand{\ms}{m\,s$^{-1}$}
\newcommand{\vs}{$v\sin{i}$}
\newcommand{\lgg}{$\log{g}$}

\newcommand{\teff}{$T_{\rm eff}$}

\newcommand{\kic}{\object{KIC 7177553}}

\newcommand{\vt}{$v_{\rm turb}$}

\slugcomment{Accepted by ApJ}

\shorttitle{The Quadruple System KIC 7177553}
\shortauthors{Lehmann et al.}

\begin{document}

\title{KIC 7177553: a quadruple system of two close binaries}

\author{H. Lehmann\altaffilmark{1},
T. Borkovits\altaffilmark{2},
S. A. Rappaport\altaffilmark{3},
H. Ngo\altaffilmark{4},
D. Mawet\altaffilmark{5},
Sz. Csizmadia\altaffilmark{6}, and
E. Forg\'acs-Dajka\altaffilmark{7}}

\altaffiltext{1}{Th\"uringer Landessternwarte Tautenburg, Sternwarte 5,\linebreak D-07778 Tautenburg, Germany; 
lehm@tls-tautenburg.de}
\altaffiltext{2}{Baja Astronomical Observatory of Szeged University, H-6500 Baja, Szegedi \'ut, Kt. 766,
Hungary; borko@electra.bajaobs.hu}
\altaffiltext{3}{Massachusetts Institute of Technology, Department of Physics, 77 Massachusetts Avenue
Cambridge, MA 02139-4307, USA; sar@mit.edu}
\altaffiltext{4}{California Institute of Technology, Division of Geological and Planetary Sciences,
1200 E California Blvd MC 150-21, Pasadena, CA 91125, USA; hngo@caltech.edu}
\altaffiltext{5}{California Institute of Technology, Astronomy Dept. MC 249-17, 1200 E. California Blvd., Pasadena, CA 91125,
USA; dmawet@astro.caltech.edu}
\altaffiltext{6}{German Aerospace Center (DLR), Institut f\"ur Planeten-forschung, Rutherfordstra\ss e 2, 12489
Berlin, Germany;\linebreak szilard.csizmadia@dlr.de}
\altaffiltext{7}{Astronomical Department, E\"otv\"os University, H-1118 Budapest, P\'azm\'any P\'eter stny.~1/A,
Hungary; e.forgacs-dajka@astro.elte.hu}

\begin{abstract}
KIC 7177553 was observed by the {\em Kepler} satellite to be an eclipsing
eccentric binary star system with an 18-day orbital period.  Recently, an
eclipse timing study of the {\em Kepler} binaries has revealed eclipse
timing variations in this object with an amplitude of $\sim$100 sec, and
an outer period of 529 days.  The implied mass of the third body is that
of a superJupiter, but below the mass of a brown dwarf.  We therefore
embarked on a radial velocity study of this binary to determine its
system configuration and to check the hypothesis that it hosts a giant
planet.  From the radial velocity measurements, it became immediately obvious that the
same {\em Kepler} target contains another eccentric binary, this one
with a 16.5-day orbital period.  
Direct imaging using adaptive optics reveals that the two binaries are
separated by 0.4$''$ ($\sim$167 AU), and have nearly the same magnitude
(to within 2\%).
The close angular proximity of the two
binaries, and very similar $\gamma$ velocities, strongly suggest that KIC
7177553 is one of the rare SB4 systems consisting of two eccentric
binaries where at least one system is eclipsing.  Both systems consist of
slowly rotating, non-evolved, solar-like stars of comparable masses.
 From the orbital separation and the small difference in $\gamma$
velocity,  we infer that the period of the outer orbit most likely lies in the range 
1000 to 3000 years. New images taken over the next few years, as well as the 
high-precision astrometry of the Gaia satellite mission, will allow us to set much 
narrower constraints on the system geometry. Finally, we note that the observed 
eclipse timing variations in the {\em Kepler} data cannot be produced by the second
binary. Further spectroscopic observations on a longer time scale will be
required to prove the existence of the massive planet.
\end{abstract}


\section{Introduction}

 Most of our knowledge of stellar masses comes from the investigation of binary stars. In particular eclipsing
binaries (EBs) can provide the absolute masses of their components. Precise absolute masses of stars across the whole
Hertzsprung-Russell diagram are urgently needed for 
testing the theories of stellar structure and evolution and, recently, more and more for establishing accurate scaling relations
for the pulsation properties of oscillating stars in the rapidly developing field of asteroseismology 
\citep[e.g.][]{2015AN....336..477A}. This will be especially important for the scaling of measured planet-star mass ratios
in such future space projects like PLATO \citep{2014ExA....38..249R} where the stellar masses will be asteroseismically determined. 

The {\it Kepler} space telescope mission \citep{2011ApJ...728..117B}, designed and launched for searching for transiting planets,
also delivered light curves of unprecedented photometric quality of a large number of EBs 
\citep[e.g.][]{2011AJ....142..160S}. Among these EBs, 
multiple systems were also found, including triply eclipsing hierarchical triples such as \object{KOI-126} \citep{2011Sci...331..562C},
\object{HD 181068} \citep{2011Sci...332..216D}, and \object{KIC 4247791}, a SB4 system consisting of two eclipsing binaries
\citep{2012A&A...541A.105L}. Such unusual objects (from the observational side) additionally offer the possibility 
for studying the tidal interaction with the third body or between the binaries in the case of quadruple systems.

EBs in eccentric orbits, on the other hand, allow for the observation of apsidal motion as a probe of stellar
interiors \citep[e.g.,][]{1993A&A...277..487C}. Moreover, such types of EBs are interesting for searching for 
tidally induced oscillations occurring as high-frequency p-modes in the convective envelopes of the components
\citep[e.g.,][]{2013MNRAS.429.2425F} or as the result of a resonance between the dynamic tides and one or
more free low-frequency g-modes as observed in \object{KOI 54} \citep{ 2011ApJS..197....4W}.
Tidally induced pulsations allow us to study tidal interaction that impacts the orbital evolution as well as
the interiors of the involved stars.

\kic\ (\object{TYC 3127-167-1}) is listed in the catalog Detection of Potential Transit Signals in the First Three Quarters of 
Kepler Mission Data \citep{2012ApJS..199...24T} with an orbital period of 18\,d.
\citet{2014MNRAS.437.3473A} combined the Kepler Eclipsing Binary Catalog 
\citep{2011AJ....141...83P,2011AJ....142..160S} with information from the Howell-Everett
\citep[HES,][]{2012PASP..124..316E}, Kepler INT \citep[KIS,][]{ 2012AJ....144...24G} and 2MASS 
\citep{ 2006AJ....131.1163S}
photometric surveys to produce a catalog of spectral energy distributions of Kepler EBs.
They estimate the temperatures of the primary and secondary components of \kic\ to be 5911$\pm$360\,K and
5714$\pm$552\,K, respectively, and derive $R_2/R_1$\,=\,0.89$\pm$0.28 for the ratio of the radii of the components.

In a recent survey of eclipse timing variations (ETVs) of EBs in the original
{\it Kepler} field {\citep[][B15 hereafter]{2015arXiv151008272B} low 
amplitude ($\sim$\,90\,sec) periodic ETVs were found and interpreted as the consequence of dynamical perturbations  
by a giant planet that revolves around the eclipsing binary in an eccentric, inclined,  1.45\,year orbit.
In order to confirm or reject the  hypothesis of the presence of a non-transiting circumbinary
planet in the  \kic\ system, we carried out spectroscopic follow-up observations.
 This article describes the analysis of these observations that led to an unexpected finding,
 namely that \kic\ is a SB4 star consisting of two binaries. This fact makes the star an extraordinary object.
As a possible quadruple system its properties  are important for our understanding of the formation and evolution of
multiple systems \citep[e.g.][]{2014prpl.conf..267R} and may be very important for the theory of planet
formation in multiple systems \citep[e.g.][]{2014sf2a.conf..135D} if one of the binaries in \kic\ hosts the suspected
giant planet.

\vspace{6mm}
\section{Spectroscopic Investigation}
	
\subsection{Observations and Data Reduction}

New spectra were taken in May to July 2015 with the Coude-Echelle spectrograph attached to the 2-m Alfred Jensch Telescope at the
Th\"uringer Landessternwarte Tautenburg. The spectra have a resolving power of 30\,000 and cover the wavelength range from 454 
to 754\,nm. The exposure time was 40 min,  allowing for the $V$\,=\,11\fm 5 star for a typical signal-to-noise ratio (SN hereafter)
of the spectra of about 100. 
The dates of observation are listed in Table\,\ref{Obs} in the Appendix together with the measured radial velocities 
(RVs hereafter, see Sect.\,\ref{RVs}).

The spectrum reduction was done using standard ESO-MIDAS packages. It included filtering of cosmic rays, bias and straylight 
subtraction, optimum order extraction, flat fielding using a halogen lamp, normalization to the local continuum, wavelength calibration using a ThAr lamp,
and merging of the Echelle orders.
The spectra were corrected for small instrumental shifts using a larger number of telluric O$_2$ lines.

\vspace{2mm}	
\subsection{Radial Velocities and Orbital Solutions}\label{RVs}				

\begin{figure*}\centering
\epsfig{figure=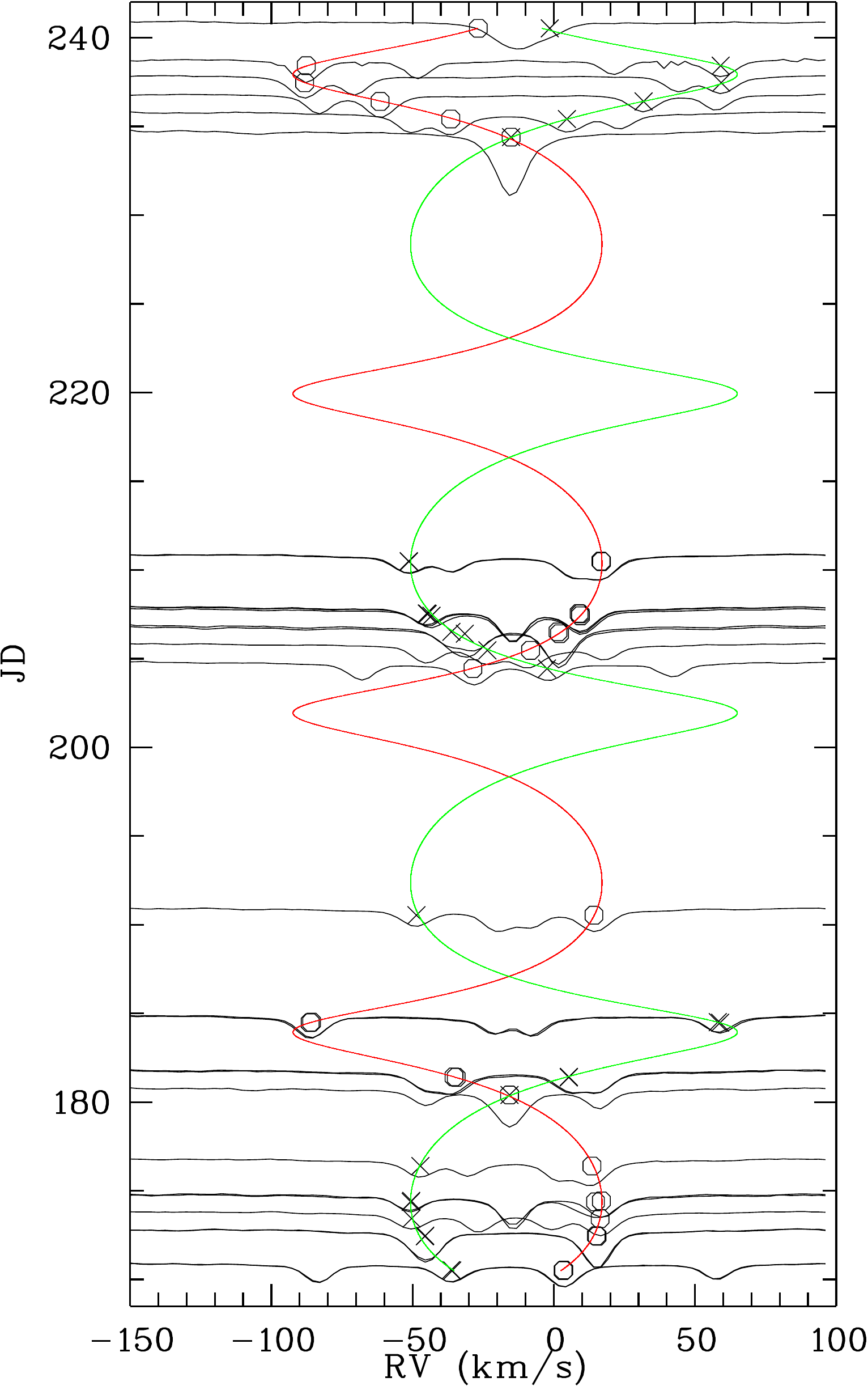, width=5.5cm, clip=}\hspace{5mm}
\epsfig{figure=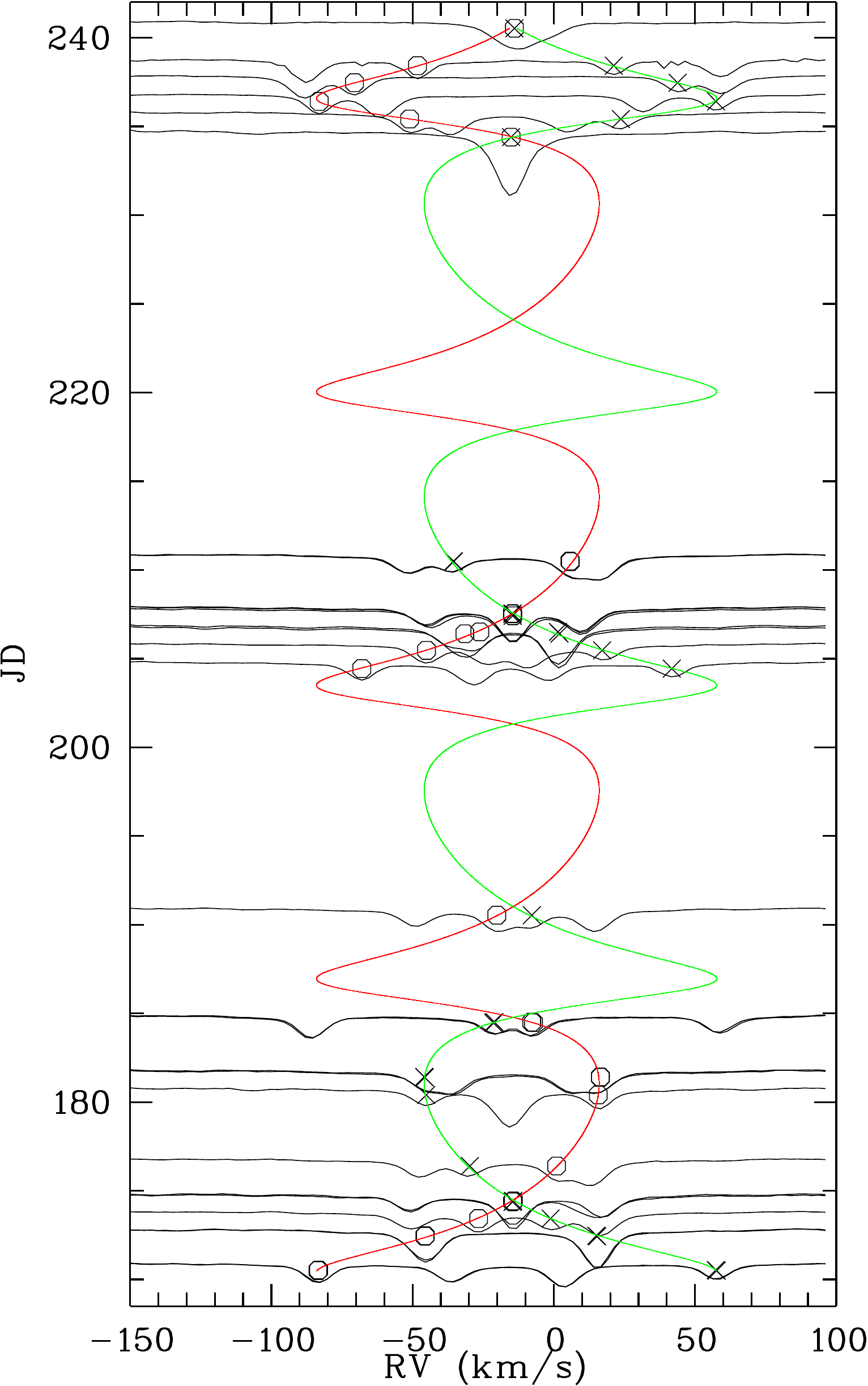, width=5.5cm, clip=}
\caption{ LSD profiles calculated from the 29 spectra, vertically arranged according to the JD (2\,457\,000+) of observation.
Measured RVs are marked by open circles and crosses, in the left panel for components $C_1$ and $C_2$ and 
in the right panel for $C_3$ and $C_4$, respectively. The red and green curves illustrate the corresponding,
calculated orbital solutions.}\label{LSD}
\end{figure*}

\begin{figure*}\centering
\epsfig{figure=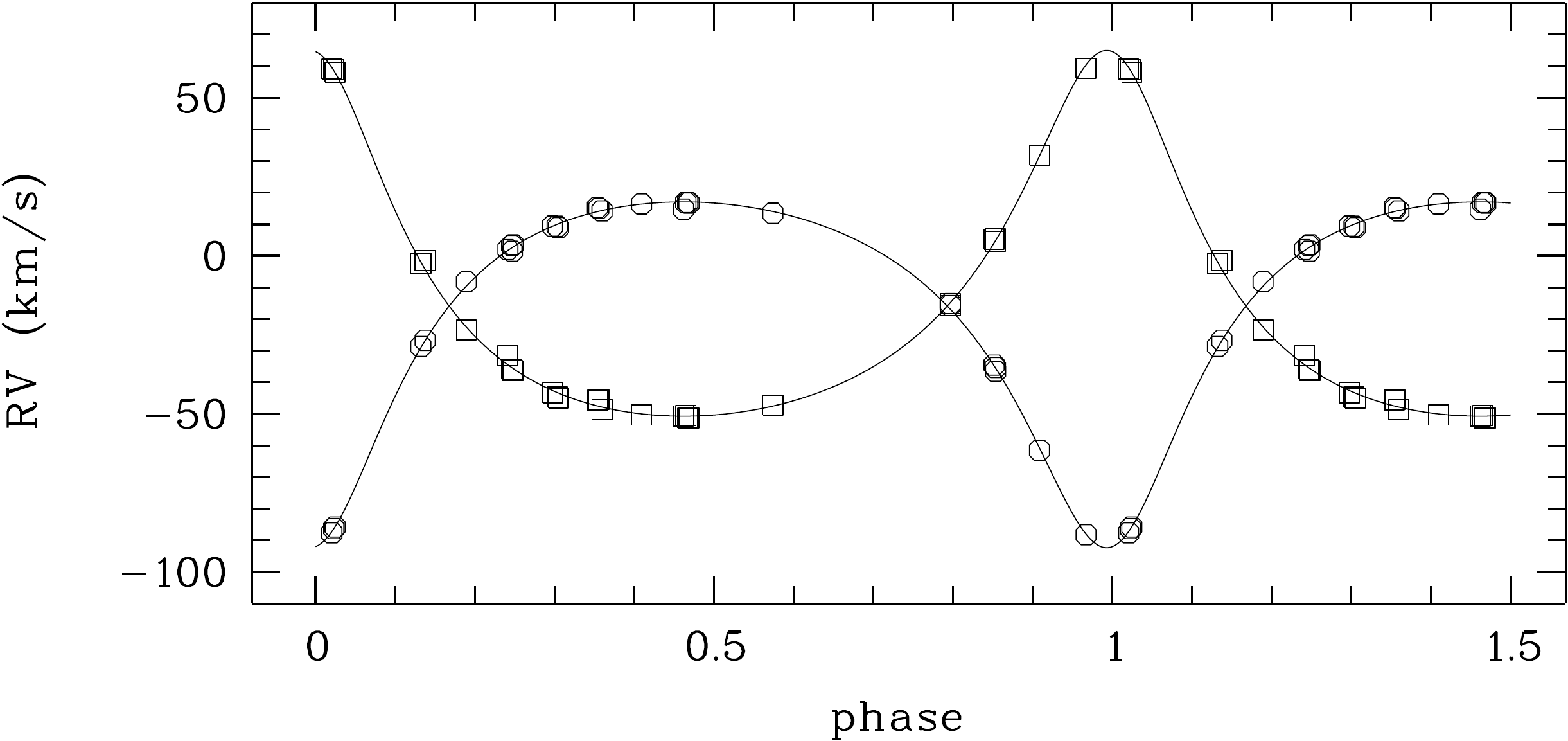, angle=-90, width=8cm, clip=}\hspace{2mm}
\epsfig{figure=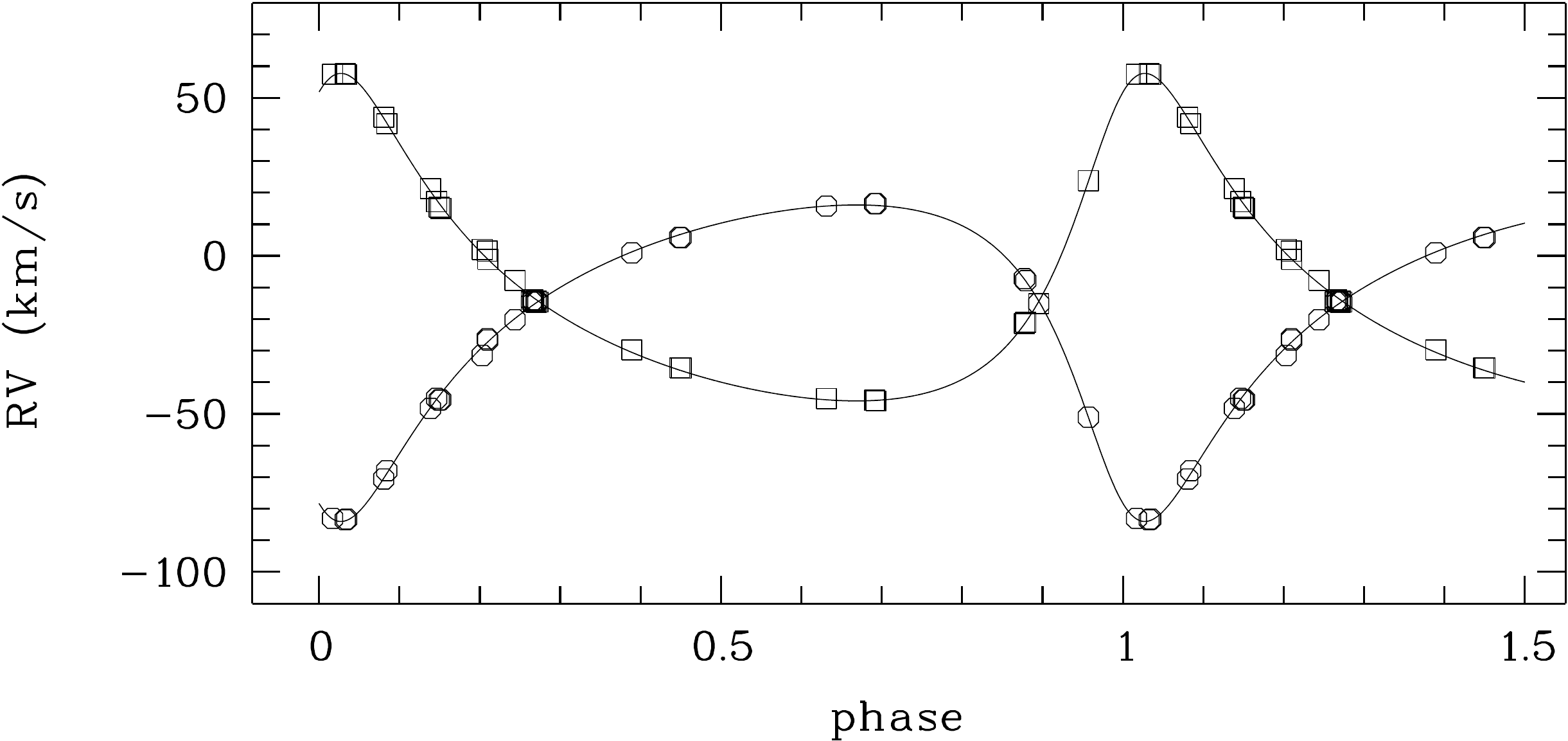, angle=-90, width=8cm, clip=}
\caption{RVs and calculated orbital curves folded with the orbital periods. 
Left: $C_1$ (open circles) and $C_2$ (open squares). 
Right: $C_3$ (open circles) and $C_4$ (open squares).
Phase zero corresponds to the time of periastron passage.}\label{OrbSolv}
\end{figure*}	
	
\begin{figure*}\centering
\epsfig{figure=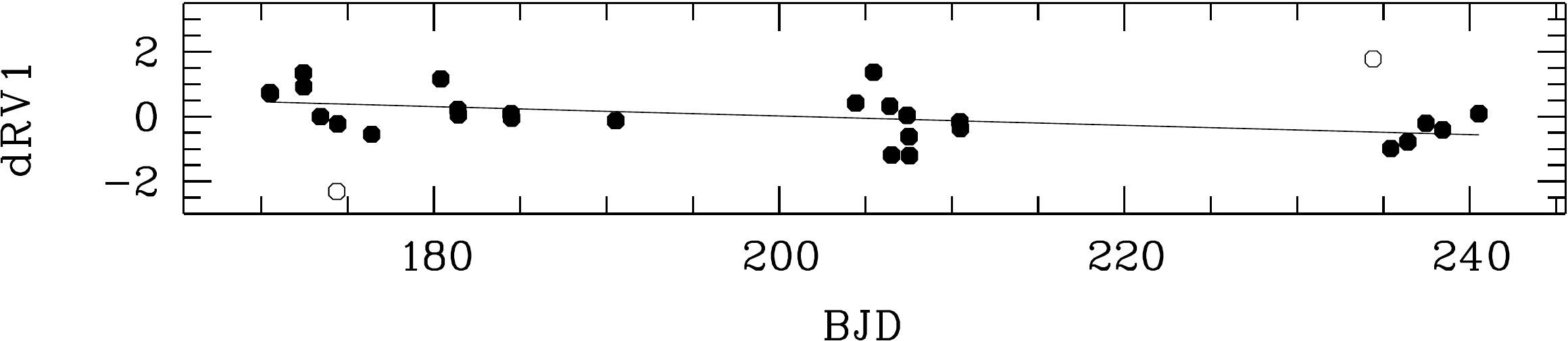, angle=-90, width=8cm, clip=}\hspace{3mm}
\epsfig{figure=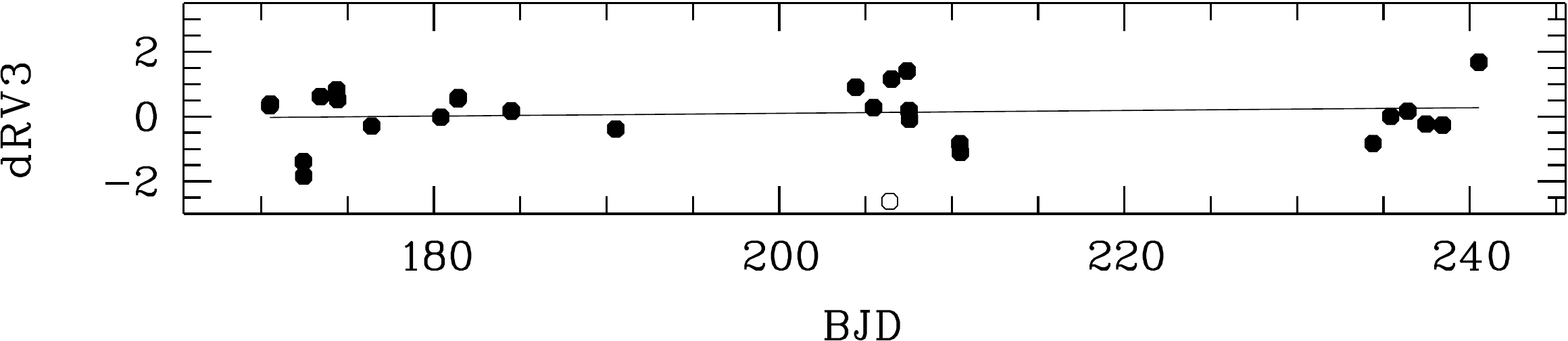, angle=-90, width=8cm, clip=}\vspace{-6mm}\\
\epsfig{figure=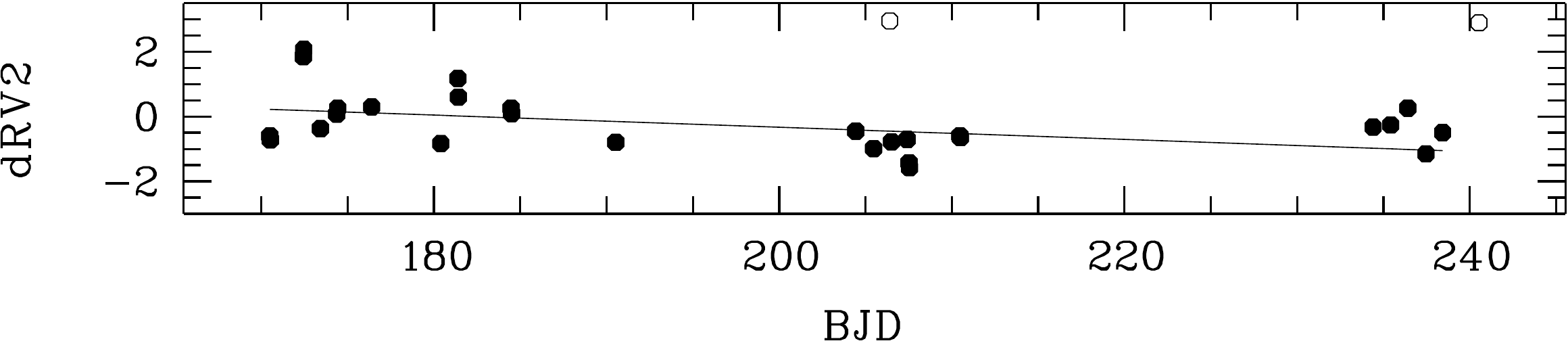, angle=-90, width=8cm, clip=}\hspace{3mm}
\epsfig{figure=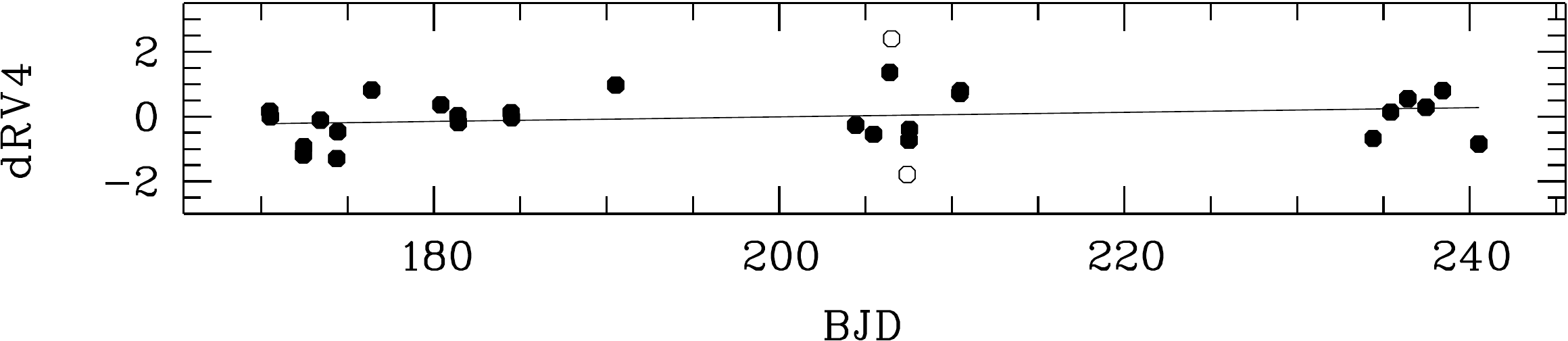, angle=-90, width=8cm, clip=}
\caption{O-C values in \kms\ versus BJD (2\,457\,000+) for components $C_1$ to $C_4$. 
The straight lines result from a linear regression calculated
from all RVs shown by filled circles. Outliers are indicated by open circles.}\label{Resid}
\end{figure*}	
		
In a first step, we used the cross-correlation of the observed spectra with an unbroadened 
synthetic template spectrum based on the 487 to 567 nm metal lines region (redwards of H$\beta$, where no
stronger telluric lines occur) to look for 
multiple components in the cross-correlation functions (CCFs hereafter). The template was
calculated with SynthV \citep{1996ASPC..108..198T} for \teff\ of 6200\,K, based on a model atmosphere 
calculated with LLmodels \citep{2004A&A...428..993S}. 
Surprisingly, there were two spectra where we could clearly see four components in the CCFs. In all other spectra, 
the components were more or less blended and we saw only one to three components.

Next, we used the least squares deconvolution \citep[LSD,][]{1997MNRAS.291..658D}) technique and
calculated LSD profiles with the LSD code by \citet{2013A&A...560A..37T}, 
based on the same line mask as we used for our before mentioned synthetic spectrum. 
Figure\,\ref{LSD} shows the LSD profiles vertically arranged according to the JD of observation.
The resolution is distinctly better than that of the CCFs, four components are now seen
in the majority of the LSD profiles. 
The measured RVs are marked and the calculated orbital curves plotted in Fig.\,\ref{LSD} show 
that we could assign all components in all LSD profiles to 
four different stars located in two different binary systems, $S_{1,2}$ and $S_{3,4}$.

The RVs of the four components, $C_1$ to $C_4$, were determined by fitting the LSD profiles with multiple Gaussians.
The mean (internal) errors of the fit were 0.25, 0.36, 0.28, and 0.37\,\kms\ for the RVs of $C_1$ to $C_4$, respectively.
We used the method of differential corrections \citep{ 1910PAllO...1...33S} to determine the orbits from the RVs.
Because of the short time span of the observations, this could be done for the two systems separately, 
without accounting for any interaction between them or for light-travel-time effects. 
In the case of $S_{1,2}$, we fixed the orbital period to the value 
known from the {\it Kepler} light-curve analysis. 			
The RVs of the four components of \kic\ determined from the multi-Gaussian fit to the LSD profiles
are listed in Table\,\ref{Obs} in the Appendix.

Figure\,\ref{OrbSolv} shows the RVs and orbital curves folded with the orbital periods.
Table\,\ref{Orbits} lists the derived orbital parameters.
The times of observation of all spectra are based on UTC. To be consistent with the {\it Kepler} DCT time scale and the results listed in 
Table\,\ref{Tab:syntheticfit}, we added 68\,sec to the calculated times of periastron passage $T$. 
 
\begin{table*}
\tabcolsep 2.7mm
\caption{Orbital solutions obtained from multi-Gaussian fits of the LSD profiles and with KOREL.}\label{Orbits}
\begin{tabular}{|l|cccc|cccc|}
\hline\hline
                      & \multicolumn{4}{c|}{LSD} & \multicolumn{4}{c|}{KOREL}\\
                      & \multicolumn{2}{c}{$S_{1,2}$}		& \multicolumn{2}{c|}{$S_{3,4}$}	   & \multicolumn{2}{c}{$S_{1,2}$}	     & \multicolumn{2}{c|}{$S_{3,4}$}	  \\
\hline
$P$ (d)               & \multicolumn{2}{c}{17.996467(17)}	& \multicolumn{2}{c|}{16.5416(70)}	   & \multicolumn{2}{c}{17.996467(17)}       & \multicolumn{2}{c|}{16.5490(38)}     \\
$e$	              & \multicolumn{2}{c}{0.3984(31)}  	& \multicolumn{2}{c|}{0.4437(35)}	   & \multicolumn{2}{c}{0.4008(17)}	     & \multicolumn{2}{c|}{0.4421(12)}       \\
$\omega\,(^\circ)$    & \multicolumn{2}{c}{6.51(63)}		& \multicolumn{2}{c|}{332.30(75)}	   & \multicolumn{2}{c}{6.76(42)}	     & \multicolumn{2}{c|}{331.17(27)}        \\  
$T$ (BJD)             & \multicolumn{2}{c}{2\,457\,184.067(27)} & \multicolumn{2}{c|}{2\,457\,186.505(34)} & \multicolumn{2}{c}{2\,457\,184.039(21)} &\multicolumn{2}{c|}{2\,457\,186.457(13)}\\
$q$                   & \multicolumn{2}{c}{0.9457(71)}  	& \multicolumn{2}{c|}{0.9664(78)}	   & \multicolumn{2}{c}{0.9359(36)}	     & \multicolumn{2}{c|}{0.9612(27)}		\\
\hline
                      & $C_1$	   & $C_2$	& $C_3$      & $C_4$	 & $C_1$      & $C_2$	   & $C_3$       & $C_4$  \\
\hline
$K$~~~(\kms)          &    54.70(26) &    57.84(34) &    50.06(30) &   51.80(28) &  54.94(17) &  58.70(13) &  50.148(94) & 52.17(11)\\ 
$\gamma$~~~~(\kms)    & $-$15.98(17) & $-$15.77(22) & $-$14.30(19) &$-$14.43(18) &	      & 	   &	         &           \\ 
rms (\kms)            &     0.873    &     1.17	    &     0.936    &    0.871	 &   0.45     &   0.35     &   0.26      &  0.30      \\
\hline
$M_{\rm min}$ (\Msun) &     1.054(14)&     0.997(12)&     0.663(9) &    0.641(9) & 1.0871(64) & 1.0174(71) & 0.6758(34)  & 0.6496(31)   \\
\hline
\end{tabular}
\tablecomments{The table lists period $P$ and orbital elements (eccentricity $e$, argument of periastron $\omega$, time of periastron passage $T$, and
mass ratio $q$) of the two systems, and the RV semi-amplitudes $K$ and individual systemic velocities $\gamma$. Errors are given in units 
of the last two digits in parentheses. The last two rows list the rms of the residuals after subtracting the orbital solution from 
the RVs and the projected masses calculated from the spectroscopic mass function.}
\end{table*}
 
The O-C residuals after subtracting the orbital solutions from the RVs are shown in Fig.\,\ref{Resid},
together with straight lines resulting from a linear regression using 2\,$\sigma$-clipping to reject outliers. 
The slopes of the regression lines are 
$-5.3\pm1.5$, $-6.9\pm2.0$, $1.6\pm1.8$, and $2.6\pm1.5$\,\kms\,yr$^{-1}$ for $C_1$ to 
$C_4$, respectively. 
These slopes describe a change in the systemic velocities of the two binaries or additional RV components of the 
single objects not included in our Keplerian orbital solutions. They are different from zero by 
3.5 times the $1\,\sigma$ error bars for $C_1$ and $C_2$ but much less or not significantly different from zero for 
$C_3$ and $C_4$. There are no outliers anymore when using 3\,$\sigma$-clipping for the linear regression,
however, and all slopes turn out to be non-significant. 

The typical accuracy in RV that we can reach with our spectrograph and reduction methods (without using an iodine cell)
for a single-lined, solar-type star with such sharp lines and spectra with SN of 100 is of about 150\,\ms. 
The rms as listed in Table\,\ref{Orbits} is distinctly higher. To check for periodic signals possibly hidden in the O-C residuals,
we performed a frequency search using the Period04
program \citep{2005CoAst.146...53L}. It did not reveal any periodicity in the residuals of any of the components.
We assume that the higher rms results from the fact that we are dealing with an SB4 star using multi-Gaussian fits
to disentangle the blended components in the LSD profiles and to derive the RVs
and assume that the listed rms stand for the measurement errors.

\subsection{Spectrum Decomposition}

We used the KOREL program \citep{1995A&AS..114..393H,Hadrava2006Ap&SS.304..337H} provided by the 
VO-KOREL\footnote{https://stelweb.asu.cas.cz/vo-korel/} web service 
\citep{2010ASPC..435...71S} for decomposing the spectra. 
Allowing for timely variable line strengths of all four components,
we got smooth decomposed spectra with only slight undulations in the single continua as they are typical for Fourier
transform-based methods of spectral disentangling \citep[see e.g.][]{2010ASPC..435..207P}. These undulations were removed by
comparing the KOREL output spectra with the mean composite spectrum and applying continuum corrections based on spline fits.

The resulting orbital parameters are listed and compared to those obtained in Sect.\,\ref{RVs} 
in Table\,\ref{Orbits}. The Fourier transform-based KOREL program
does not deliver the systemic velocity and also does not provide the errors of the derived parameters. 
The parameter errors were calculated by solving the orbits with the method of differential
corrections using the orbital parameters and line shifts delivered by the KOREL program as input. 
Comparing the results with those obtained from the LSD-based RVs, we see that there is agreement within the 1$\sigma$
error bars. The rms of the residuals of the KOREL solutions for the single components is distinctly lower, however.
The last row lists the projected masses calculated from the spectroscopic mass functions.
It can be seen that the minimum masses derived for system $S_{3,4}$ are distinctly lower than those of $S_{1,2}$ which implies
different viewing angles for the two systems.

\subsection{Spectrum Analysis}

We used the spectrum synthesis-based method as described in \citet{2011A&A...526A.124L}   
to analyze the decomposed spectra of the four components. The method compares the observed spectra with synthetic ones
using a huge grid in stellar parameters. A description of an advanced version of the program can also be 
found in \citet{2015A&A...581A.129T}. Synthetic spectra are computed with SynthV 
\citep{1996ASPC..108..198T} based on atmosphere models calculated with LLmodels \citep{2004A&A...428..993S}.
Both programs consider plan-parallel atmospheres and work in the LTE regime. 
Atomic and molecular data were taken from the VALD\footnote{http://vald.astro.univie.ac.at/~vald/php/vald.php} 
data base \citep{2000BaltA...9..590K}.

One main problem in the spectrum analysis of multiple systems is that programs for spectral disentangling like KOREL deliver 
the decomposed spectra normalized to the common continuum of all involved stars. To be able to renormalize the spectra to the continua 
of the single stars, we have to know the continuum flux ratios between the stars in the considered wavelength range. 
These flux ratios can be obtained during the spectrum analysis itself from a least squares fit between the
observed and the synthetic spectra as we show in Appendix\,\ref{Mathe}. We extended our program accordingly and tested
the modified version successfully on synthetic spectra.

The analysis is based on the wavelength interval 455-567\,nm that includes H$\beta$ and is almost free of telluric
contributions.
It was performed using four grids of atmospheric parameters for the four stars. Each grid consists of (step widths in
parentheses) \teff\ (100\,K), \lgg\ (0.1\,dex), \vt\ (1\,\kms), \vs\ (1\,\kms), and scaled solar abundances
[M/H] (0.1\,dex). The analysis includes all four spectra simultaneously, which are coupled via the flux ratios.
To obtain the optimum flux ratios and renormalize the spectra, we solved Eqn.\,\ref{LGS} (see the Appendix) for each combination of atmospheric
parameters. 

The results of the analysis are listed in Table\,\ref{Analysis}. The given errors where obtained from $\chi^2$
statistics as described in \citet{2011A&A...526A.124L}. They where calculated from the full grid in all parameters per star, 
i.e. the errors include all interdependencies between the different parameters of one star.  We did not have enough computer power
to include the interdependencies between the parameters of different stars which interfere in the simultaneous analysis via 
the flux ratios, however.

\begin{table}\centering
\tabcolsep 1.4mm
\caption{Atmospheric parameters of the four stars.}
\begin{tabular}{lrrrr}
\hline\hline
 & \multicolumn{1}{c}{$C_1$} & \multicolumn{1}{c}{$C_2$} &\multicolumn{1}{c}{$C_3$} &\multicolumn{1}{c}{$C_4$}\\
 \hline
$[M/H]$~(dex)	 & 0.00(11)  & -0.10(13) & -0.12(13) & -0.12(13)\\
\teff~~~~~(K)	 & 5800(130) & 5700(150) & 5600(150) & 5600(140)\\
\lgg~~~~(c.g.s.) & 4.75(38)  & 4.55(40)  & 4.63(38)  & 4.59(35) \\
\vt~~~(\kms)	 & 1.76(60)  & 1.23(63)  & 1.29(65)  & 1.02(59) \\
\vs~~(\kms)	 & 1.3(4.2)  & 3.9(3.7)  & 5.8(3.6)  & 2.5(4.2) \\
$f$     	 & 0.30 & 0.23 & 0.24 & 0.23\\ 
\hline
\end{tabular}\label{Analysis}
\tablecomments{Errors are given in parentheses, in units of the last digits. $f$ is the continuum flux ratio of the components.}
\end{table}	  

\section{Light-Curve Analysis}

\subsection{Long-cadence data}

For the photometric light-curve analysis we downloaded the preprocessed, full $Q0-Q17$
{\it Kepler} long cadence (LC) data series from the Villanova site\footnote{http://keplerebs.villanova.edu/} of the {\it Kepler} 
Eclipsing Binary Catalog \citep{2011AJ....141...83P,2011AJ....142..160S,2012AJ....143..123M}. 
Note, the same data set was used for the ETV analysis of \kic\, which is described in detail in B15.
 This $\sim$1470\,d-long light curve was folded, binned and averaged for the analysis. 
The out-of-eclipse sections were binned and averaged equally into $0\fp002$-phase-length cells,
while for the narrow primary and secondary eclipses, i.e., in the ranges of $\phi_\mathrm{pri}$\,=\,$[-0\fp005;0\fp005]$ 
and $\phi_\mathrm{sec}$\,=\,$[0\fp737;0\fp747]$, a four-times denser binning and averaging was applied.
The resulting folded light curve is shown in Fig.\,\ref{prisecfit}.

\begin{figure}\centering
\epsfig{figure=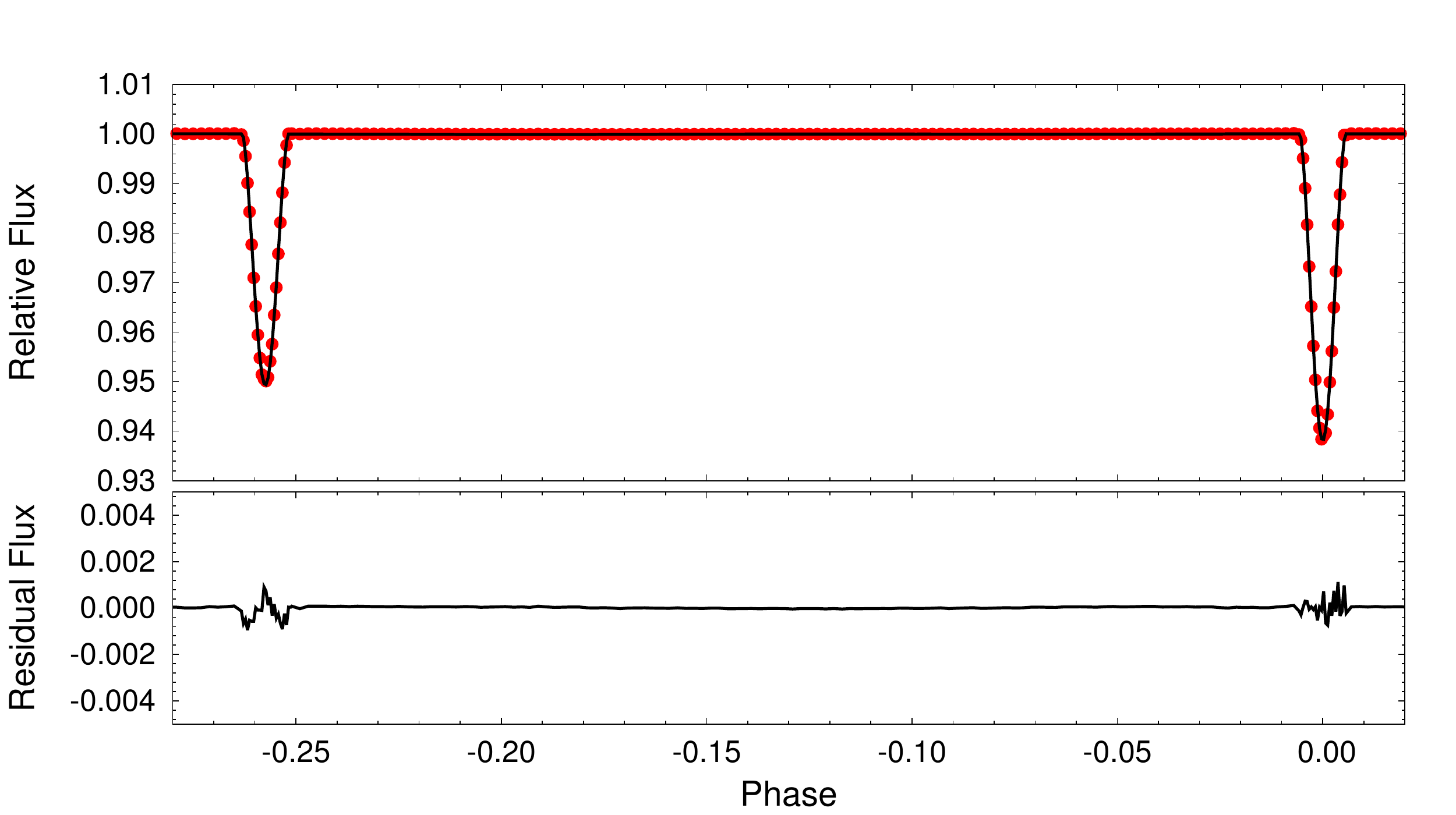, width=8.8cm, clip=}
\caption{{\it Kepler} light curve of KIC\,7177553. {\it Upper panel:} Folded, binned, averaged light curve (red circles) 
together with the model solution (black line). {\it Lower panel:} Residual light curve.}
\label{prisecfit}
\end{figure}

The light-curve analysis was carried out with the {\sc Lightcurvefactory} program
\citep{2013MNRAS.428.1656B,2014MNRAS.443.3068B}.  The primary fitted parameters 
were the initial epoch $T_0$, orbital eccentricity ($e$), argument of periastron ($\omega$), inclination ($i$), 
fractional radii of the stars ($r_{1,2}=R_{1,2}/a$), effective temperature of the secondary ($T_2$), 
luminosity of the primary ($L_1$), and the amount of third light ($l_3$). The effective temperature of the primary 
($T_1$) and the mass ratio ($q$), as well as the chemical abundances, were taken from the spectroscopic 
solution. The other atmospheric parameters, such as limb darkening (LD), gravity brightening coefficients, and 
bolometric albedo were set in accordance with the spectroscopic results and  also kept fixed. For LD, 
the logarithmic law \citep{1970AJ.....75..175K} was applied, and the coefficients were calculated 
according to the passband-dependent precomputed tables\footnote{http://phoebe-project.org/1.0/?q=node/110} 
of the {\sc PHOEBE} team \citep{2005ApJ...628..426P,2011ascl.soft06002P}
which  are based on the tables of \citet{2004astro.ph..5087C}.  Table~\ref{Tab:syntheticfit} lists the 
parameters obtained from the light curve solution, together with some other quantities which can be calculated 
by combining the photometric and spectroscopic results.
Figure\,\ref{prisecfit} compares our model solution with the observed light curve.

The out-of-eclipse behavior of the light curve merits some further discussion. The primary and secondary
eclipse depths are 60\,000 ppm and 50\,000 ppm, respectively. By contrast, all of the physical out-of-eclipse effects, 
both expected and observed, are $\lesssim 60$ ppm. 
From a simple Fourier series we found several residual sinusoidal features in the folded light curve at a number 
of frequencies near to higher harmonics of the orbit.  The amplitudes of these sinusoids ranged from 10 to 37 ppm. 
A comparison with the amplitudes expected from different physical effects gives the following picture,
where our estimations are based on \citet{2011ApJ...728..139C}:
The asymmetric ellipsoidal light variation amplitudes are $\sim$5 ppm and $\sim$60 ppm at apastron and periastron, respectively.  
The Doppler boosting (DB) effect has a peak value of $\sim$300 ppm for each one of the stars, but their velocities are 180$^\circ$
out of phase.  Since the luminosity of the two stars differs by only $\sim$3\% (see Table\,\ref{Tab:syntheticfit}), 
this leads to a net DB amplitude of not more than 10 ppm.  The illumination (or reflection effect) ranges between 
$\sim$4 ppm at apastron to $\sim$30 ppm at periastron, after taking into account that the first order terms at the 
orbital frequency essentially cancel because of the twin nature of the two stars.  

\begin{table}\centering
\tabcolsep 1mm
\caption{Orbital and stellar parameters of the $S_{1,2}$ system.} 
 \label{Tab:syntheticfit}
 \begin{tabular}{@{}lrclrcl}
  \hline\hline
\multicolumn{7}{c}{orbital parameters and third light contribution} \\
\hline
 $P_\mathrm{orb}$ (d)& 17.996467    &$\pm$&0.000017\\
 $T$                 & 5\,690.213   &$\pm$&0.012 \\
 $T_\mathrm{MIN I}$  & 4\,954.545842&$\pm$&0.00020\\
 $a$ (R$_\odot$)     & 36.76	    &$\pm$&0.15\\
 $e$                 & 0.3915	    &$\pm$&0.0010\\
 $\omega$ ($\degr$)  & 3.298	    &$\pm$&0.061\\
 $i$ ($\degr$)       & 87.679	    &$\pm$&0.055\\
 $q_\mathrm{spec}$   & 0.9457	    &$\pm$&0.0071\\
 $l_3$               & 0.453	    &$\pm$&0.043\\
\hline
\multicolumn{7}{c}{Fixed coefficients and deduced stellar parameters} \\
   & \multicolumn{3}{c}{$C_1$} & \multicolumn{3}{c}{$C_2$} \\
  \hline
 $x_{\rm bol}$       & \multicolumn{3}{c}{ 0.6937} &  \multicolumn{3}{c}{ 0.6954}\\
 $y_{\rm bol}$       & \multicolumn{3}{c}{ 0.1603} &  \multicolumn{3}{c}{ 0.1692}\\
 $x_{\rm kep}$       & \multicolumn{3}{c}{ 0.6662} &  \multicolumn{3}{c}{ 0.6700}\\
 $y_{\rm kep}$       & \multicolumn{3}{c}{ 0.1923} &  \multicolumn{3}{c}{ 0.1924}\\
 $A$                 & \multicolumn{3}{c}{ 0.50  } &  \multicolumn{3}{c}{ 0.50  }\\
 $\beta$             & \multicolumn{3}{c}{ 0.32  } &  \multicolumn{3}{c}{ 0.32  }\\
 $r$                 & 0.02556&$\pm$&0.00010 & 0.02561&$\pm$&0.00010 \\
 $M$ (M$_\odot$)     & 1.043  &$\pm$&0.014   & 0.986  &$\pm$&0.015   \\
 $R$ (R$_\odot$)     & 0.940  &$\pm$&0.005   & 0.941  &$\pm$&0.005   \\
 $T_\mathrm{eff}$ (K)& 5800   &$\pm$&130     & 5740   &$\pm$&140     \\
 $L$ (L$_\odot$)     & 0.88   &$\pm$&0.08    & 0.85   &$\pm$&0.08    \\
 $\log g$ (dex)      & 4.517  &$\pm$&0.008   & 4.491  &$\pm$&0.008   \\
  \hline
 \end{tabular}
 \tablecomments{$T$ and $T_\mathrm{MIN I}$ are BJD 2\,450\,000+, 
$l_3$ is the photometric third light contribution,  $x_{\rm bol}$, $y_{\rm bol}$ the  
linear and logarithmic bolometric LD coefficients, $x_{\rm kep}$, $y_{\rm kep}$ the linear and logarithmic
LD coefficients for the Kepler passband, $A$ the coefficient for the bolometric albedo,
$\beta$ the gravitational brightening exponent, and $r$ the fractional radius.}
\end{table}

Finally, we allowed the {\em amplitudes} of these sinusoids to be 
free parameters in the fitting procedure in a purely mathematically way, together with the physical modeling of the 
well-known ellipsoidal and other effects.  These terms have only a minor influence on the light curve solution, 
however, which is mainly based on the eclipses whose amplitudes are three orders of magnitude larger. 

As one can see  from Tables~\ref{Analysis} and \ref{Tab:syntheticfit}, the photometric solution, and 
especially the third  light contribution
to the {\it Kepler} light curve,  are in good agreement with the spectroscopic results.
It confirms that the spectrum decomposition and our derivation
of spectroscopic flux ratios, both applied to an SB4 star for the first time, give quite reliable results.

\subsection{Short-cadence data}

The combined spectroscopic-photometric analysis revealed that the \kic\ system consists of four very 
similar solar-like main sequence stars and thus we might expect to find solar-like oscillations in the Kepler 
light curve. The frequency of maximum power, $\nu_{\rm max}$, can be estimated using the scaling relation
\begin{equation}
\nu_{\rm max} = \frac{ M/M_\sun(T_{\rm eff}/T_{{\rm eff},\sun})^{3.5}}{L/L_\sun}\,\nu_{{\rm max},\sun}
\end{equation}
\citep{1991ApJ...368..599B}. From the values given in Tables\,\ref{Analysis} and \ref{Tab:syntheticfit}
we estimate that $\nu_{\rm max}$ should lie in the range 2650-3700\,$\mu$Hz, or 230-320\,d$^{-1}$, far beyond
the Nyquist frequency of the LC data of 24.469\,d$^{-1}$.
Unfortunately, only one short-cadence (SC) run spanning 30\,d exists, having a Nyquist frequency of greater
than 700\,d$^{-1}$. We clipped off the four eclipses that 
appear in this segment and ran it through a high-pass filter with a cuton frequency of 0.5\,d$^{-1}$. 
Figure\,\ref{FFT} shows the result of a Fourier transform-based frequency search. No hint to solar-like 
oscillations could be found. Only two isolated peaks corresponding to periods of 3.678 and 3.269 min appear. 
These are known artifacts of  the Kepler SC light curves representing the 7th and 8th harmonics of the inverse 
of the LC sampling of 29.4244 min \citep[see e.g.][]{2010ApJ...713L.160G}. 
No further prominent peaks or typical bump in the periodogram could be detected. 

Solar-like oscillations have been found for most of the investigated solar-type and red subgiant and giant stars 
\citep[e.g.][]{2011ApJ...742L...3W,2013ARA&A..51..353C} and, at first sight, the lack of such finding in our
data might seem surprising. There is a simple explanation, however. The pulsation amplitude strongly decreases with
increasing $\nu_{\rm max}$ \citep{2014ApJ...783..123C}, making solar-like pulsations less detectable for 
stars of higher $\log(g)$ (or easier to detect for giant than for main sequence stars). The 
detectability, on the other hand, depends on the noise background, the apparent brightness of the star, and 
the length of the observation cycle. \citet{2011ApJ...732...54C} performed an investigation of
the detectability of oscillations in solar-type stars observed by Kepler. From their Figure\,6 it can easily
be seen that our object is too faint (or the observing period too short), 
and the detection probability based on one SC run is close to zero.

\begin{figure}
\epsfig{figure=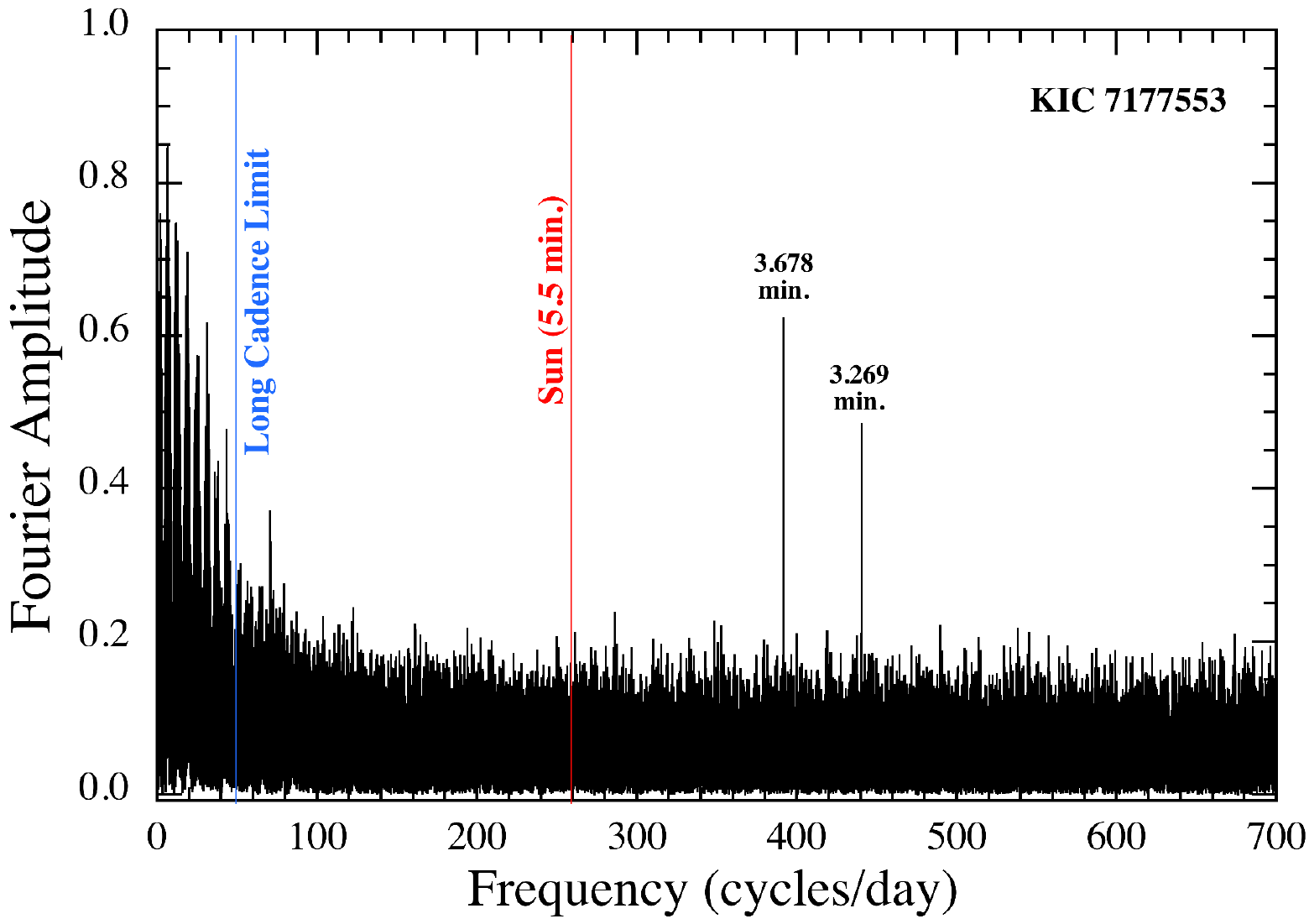, width=8.6cm, clip=}
\caption{Periodogram of the Kepler SC data of KIC\,7177553.}
\label{FFT}
\end{figure}

\section{Direct Imaging}

An examination of the UKIRT J-band image of \kic\ showed only a single stellar image.  
From the lack of any elongation of the image, we concluded that the angular separation of the two binaries is 
$\lesssim 0.5''$. To obtain better constraints, we imaged the object on 2015 October 26 UT with the NIRC2 instrument 
(PI: Keith Matthews) on Keck II using $K_s$ band (central wavelength $2.146\,\mu m$) natural guide star imaging 
with the narrow camera setting (10\,mas\,pixel$^{-1}$). To avoid NIRC2's noisier lower left quadrant, we used a 
three-point dither pattern. We obtained 6 images with a total on-sky integration time of 30 seconds. We used dome 
flat fields and dark frames to calibrate the images and also to find and remove image artifacts. 

\begin{figure}\centering
\epsfig{figure=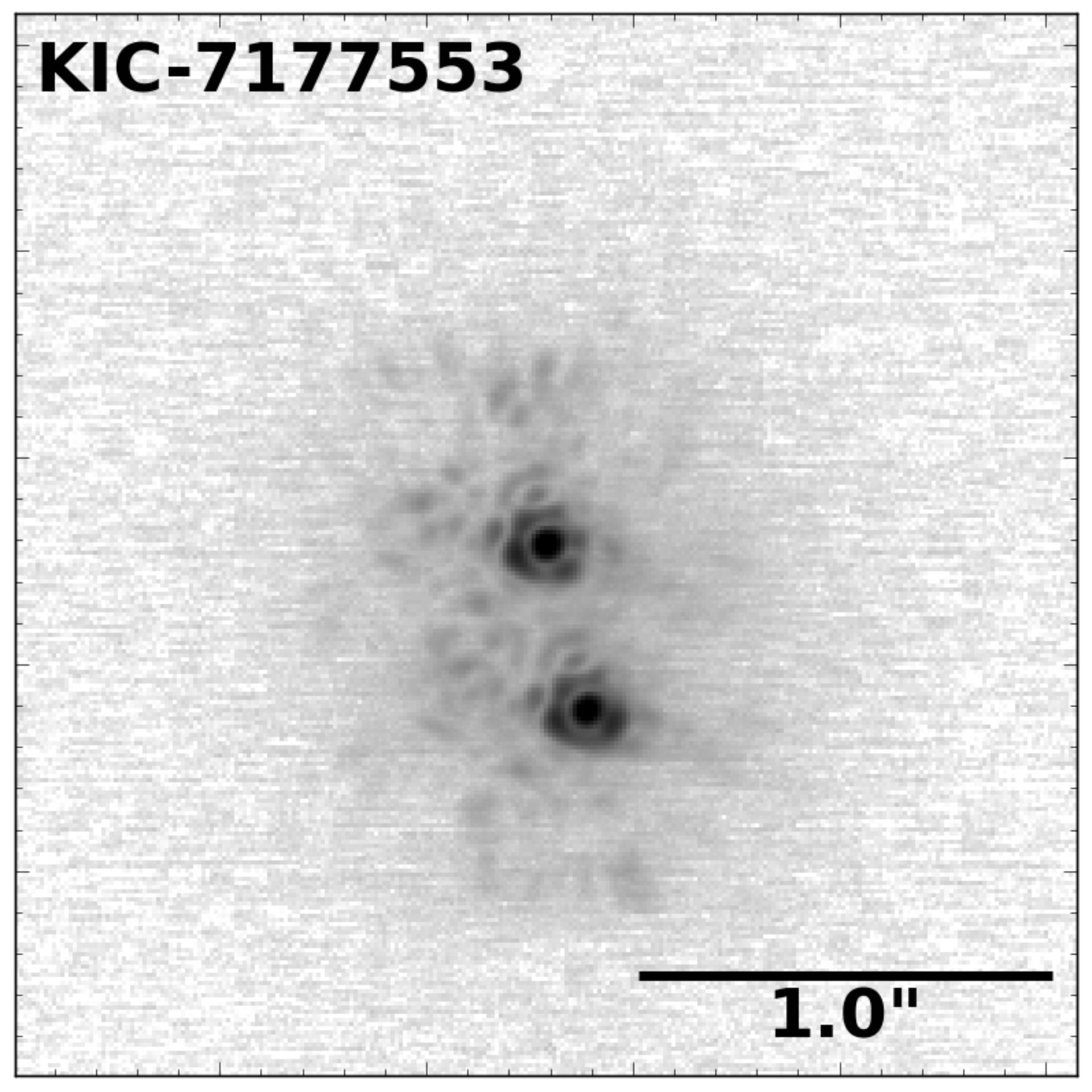, width=7.5cm, clip=}
\caption{Stacked image produced from the six NIRC2 frames showing the two well-separated binaries of KIC\,7177553. 
 The intensity is displayed on a logarithmic scale.}
\label{Keck}
\end{figure}

 The resultant stacked AO image is shown in Fig.\,\ref{Keck}, where we see two essentially twin images separated by 0.4$''$.
For each calibrated frame, we fit a two-peak PSF to measure the flux ratio 
and on-sky separations. We chose 
to model the PSF as a Moffat function with a Gaussian component. The best-fit PSF model was found over a circular 
area with a radius of 10 pixels around each star (the full width at half-maximum of the PSF was about 5 pixels).
More details of the method can be found in \citet{2015ApJ...800..138N}.

For each image, we also computed the flux ratio by integrating the best-fit PSF model over the same circular area. 
When computing the separation and position angle, we applied the astrometric corrections from 
\citet{2010ApJ...725..331Y} to account for the NIRC2 array's distortion and rotation.  Finally, we report the 
mean value and the standard error on the mean as our measured values of flux ratio (primary to secondary), 
separation, and position 
angle of the system, which are $1.018\pm0.005$, $410.4\pm1.5$\,mas, and $193.6\pm0.2$\,deg E of N, respectively.

\section{Configuration of the Quadruple}

As was mentioned in the Introduction, the B15 study found low amplitude, $\sim$\,1.45\,yr-period 
ETVs in \kic, which they interpreted as the perturbations  by a giant planet. The short period of 1.45\,yr, 
together with a total mass of both binaries of about 4\,M$_\odot$ estimated from the spectrum analysis,
would yield a separation of the two systems of only 2\,AU
and dynamically forced ETVs of the order of 1.5 hours (see Eqn.~11 in B15).
It is evident from  the results
of direct imaging that the 16.5\,d-period binary located in such a distant system and separated by $\sim$\,0.4 arcsec 
cannot produce such a signal.}

 We also conclude that there is no evidence for either light-travel-time effect or dynamical perturbations
caused by the 16.5\,d binary in the $\sim$\,4\,year-long {\it Kepler} observations of \kic.
This fact does not eliminate the possibility that the two binaries form a quadruple system, of course,
but indicates that the period of the orbital revolution of the two binaries around each other
must exceed at least a few decades.
 
Based on the quite reasonable approximation that all four stars are of spectral type G2\,V and taking the (total)
apparent magnitude, $m_V=11.629\pm0.020$, and color index, $B-V=0.741\pm0.028$, from 
\citet{2012PASP..124..316E} and $M_V=4.82\pm0.01$, $(B-V)_0=0.650\pm0.01$ from the solar values as given in 
\citet{2000asqu.book.....C},
we estimate that the distance to this quadruple system is $D=406\pm10$\,pc. In that case, the projected 
physical separation is $s=167\pm5$\,AU and the orbital period $P$ must be larger than 1000\,yr
(note that we assigned to the errors of $D$ and $s$ twice the values that would follow from error propagation, 
accounting for the approximation that all four stars have identical properties).

Because we measure only two instantaneous quantities related to the outer orbit, $s$, the separation of 
the two components, and $\Delta\gamma$, the relative radial velocity between the two binaries (or difference in
$\gamma$-velocities), we obviously cannot 
uniquely determine the orbital properties of the quadruple system.  However, we can set some quite meaningful 
constraints on the outer orbit.

Starting with the simpler circular orbit case, we can show that:
\begin{eqnarray}
s & = & a \sqrt{\cos^2{i} + \cos^2{\phi}\sin^2{i}} \\
\Delta\gamma & = & \sqrt{\frac{G M}{a}}\cos{\phi}\sin{i}
\end{eqnarray}
where  $a$ is the orbital separation, $\phi$ the orbital phase, $i$ the orbital inclination angle, and 
$M$ the total system mass.  The unknown orbital phase can be eliminated to find a cubic expression for the orbital separation:
\begin{equation}
a^3 \left(\frac{\Delta\gamma^2}{G M}\right) + a^2 \cos^2 i- s^2 =0
\end{equation}

In spite of the fact that we do not know the orbital inclination angle, $i$, we can still produce a probability 
distribution for $a$ (and hence $P_{\rm orb}$) via a Monte Carlo approach.  For each realization of the system 
we choose a random inclination angle with respect to an isotropic set of orientations of the orbital angular 
momentum vector.  In addition, because there are uncertainties in the determination of 
$s$ (accruing from the uncertainty in the distance), and in $\Delta\gamma$ (see Table\,\ref{Orbits}), we also choose specific 
realizations for these two quantities using Gaussian random errors.  In particular we take $s=167\pm5$\,AU and 
$\Delta\gamma = 1.5 \pm 0.28$\,\kms, both as 1-$\sigma$ uncertainties.  We then solve Eqn.~(3) for $a$, and we also record 
the corresponding value of $i$.  If any inclination leads to a non-physical solution of Eqn.~(3), we discard it, 
including the value of $i$ that led to it.  We repeat this process some $10^8$ times to produce our distributions. 

In a similar fashion, for each realization of the system, we can also compute the expected sky motion of the vector 
connecting the two stars.  In particular, we find:
\begin{eqnarray}
\frac{\dot s}{s} & = & -\frac{2 \pi}{P_{\rm orb}} \left(\frac{a}{s}\right)^2 \sin{\phi}\cos{\phi}\sin^2{i} \\
\dot \Theta & = & \frac{2 \pi}{P_{\rm orb}} \left(\frac{a}{s}\right)^2 \cos i
\end{eqnarray}
where $\Theta$ is commonly referred to as the ``position angle'' on the sky.

We can also find a $P_{\rm orb}$ distribution for the eccentric orbit case.  We do this by deriving equations 
analogous to (1) and (2) for the circular orbit case, except that we now must introduce two more unknown quantities, 
namely $e$, the outer eccentricity, and $\omega$, the corresponding longitude of periastron. These expressions can 
be written schematically as:
\begin{eqnarray}
s & = & a \sqrt{f_1(e,i,\omega,E)+f_2(e,i,\omega,E)} \\
\Delta\gamma & = & \sqrt{\frac{G M}{a}} g(e,i,\omega,E)
\end{eqnarray}
where $E$ is the eccentric anomaly at the time of our measurements.  The explicit expressions for $f$ and $g$ are:
\begin{eqnarray}
f_1 & = & \left[(\cos E-e)\cos \omega -\sqrt{1-e^2} \sin E \sin \omega\right]^2 \\ 
f_2 & = & \left[\sqrt{1-e^2}\sin E \cos \omega +(\cos E-e) \sin \omega \right]^2 \cos^2 i \\
g   & = & \frac{\left(\sqrt{1-e^2} \cos E \cos \omega - \sin E \sin \omega \right) \sin i}{1-e \cos E}
\end{eqnarray}

\begin{figure}\centering
\epsfig{figure=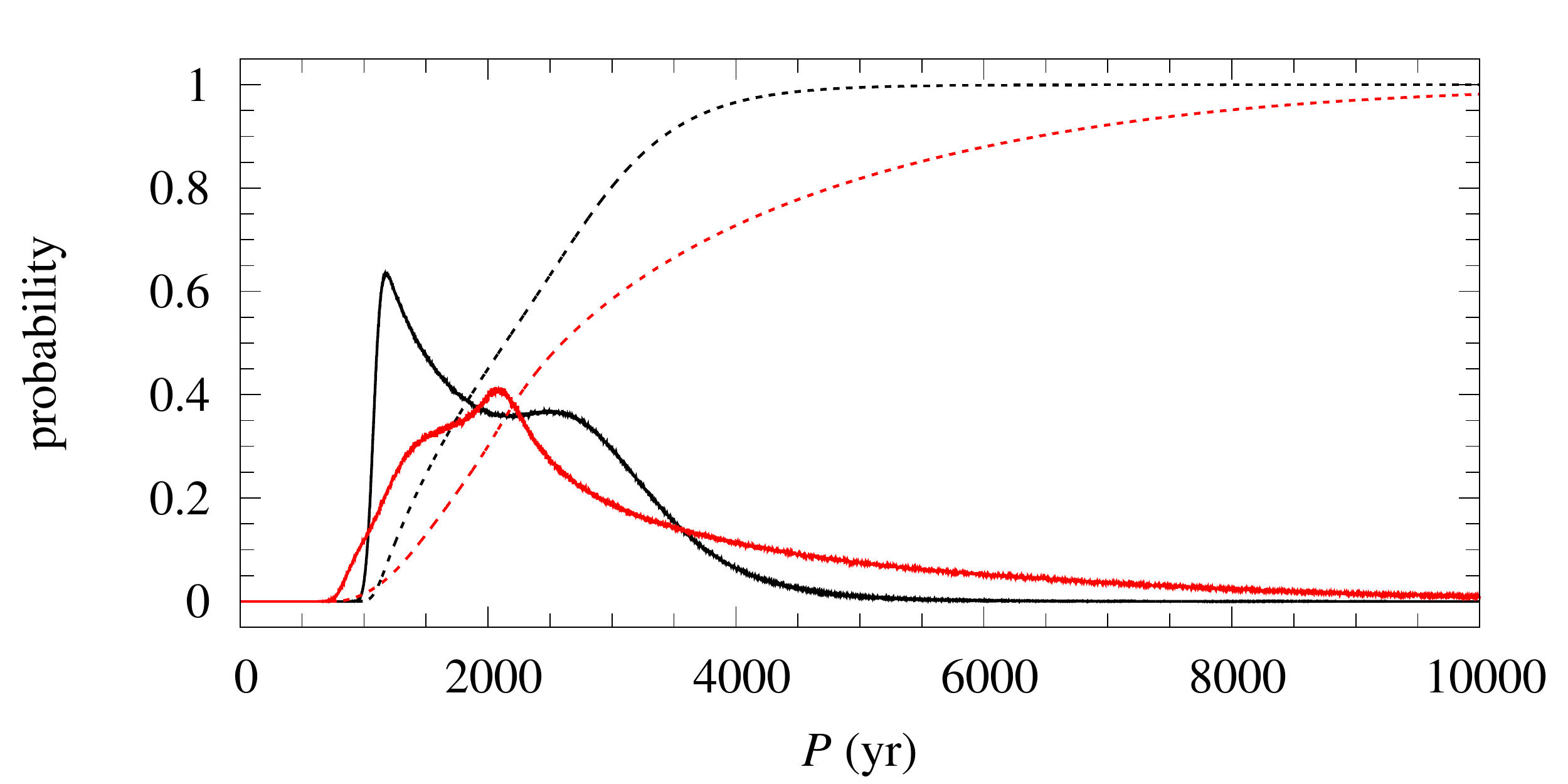, width=8.4cm, clip=}
\caption{PDFs (continuous lines) and corresponding cumulative probabilities (dashed lines) for circular (black)
and eccentric (red) orbits. For a better visualization, we scaled the PDFs by a factor of 100.}\vspace{2mm}
\label{Prob_a}
\epsfig{figure=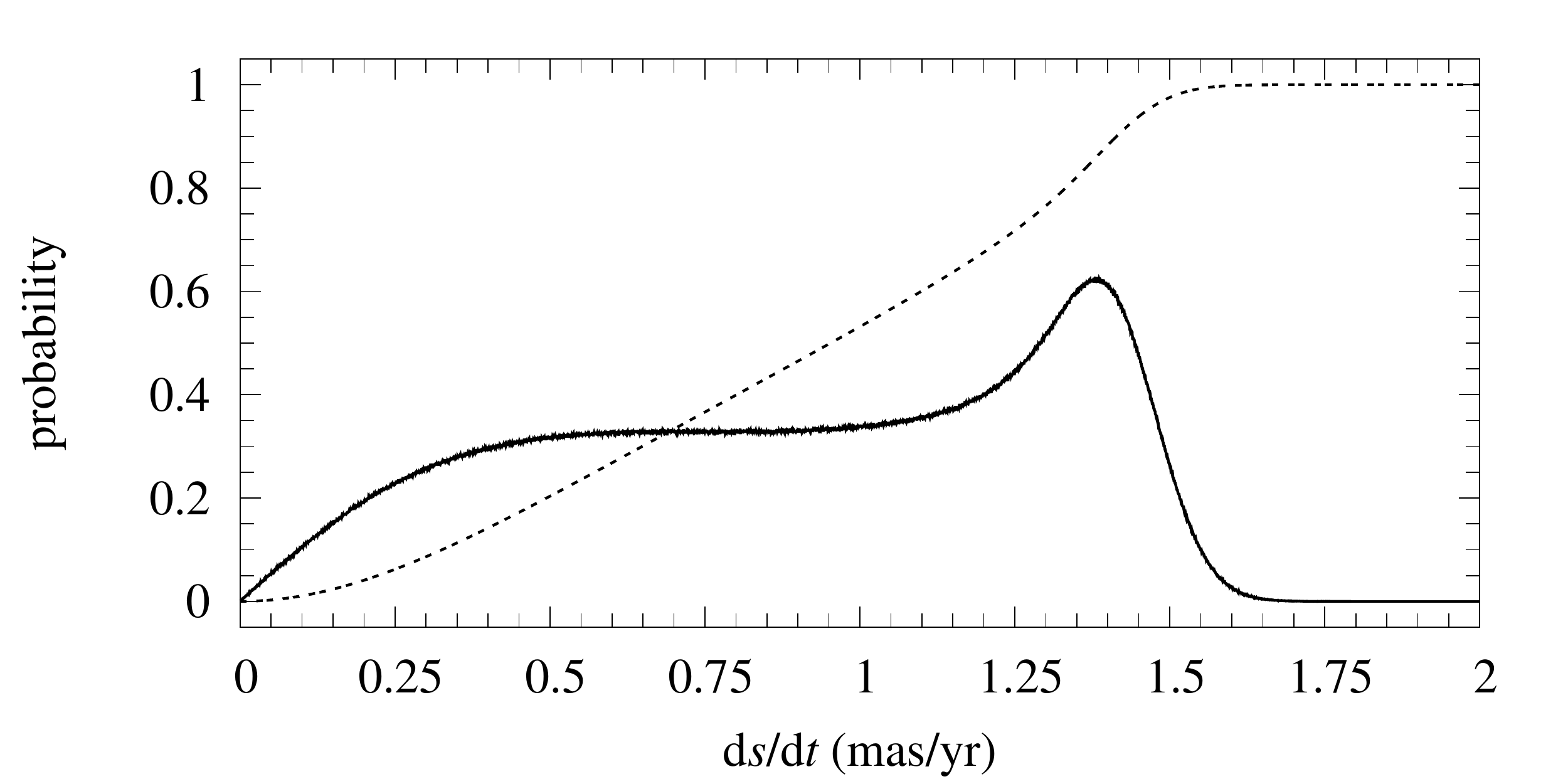, width=8.4cm, clip=}
\caption{As Fig.\,\ref{Prob_a} but for $\dot s$ and circular orbits. 
The PDF was scaled by a factor of 0.5.}\vspace{4mm}
\label{Prob_b}
\epsfig{figure=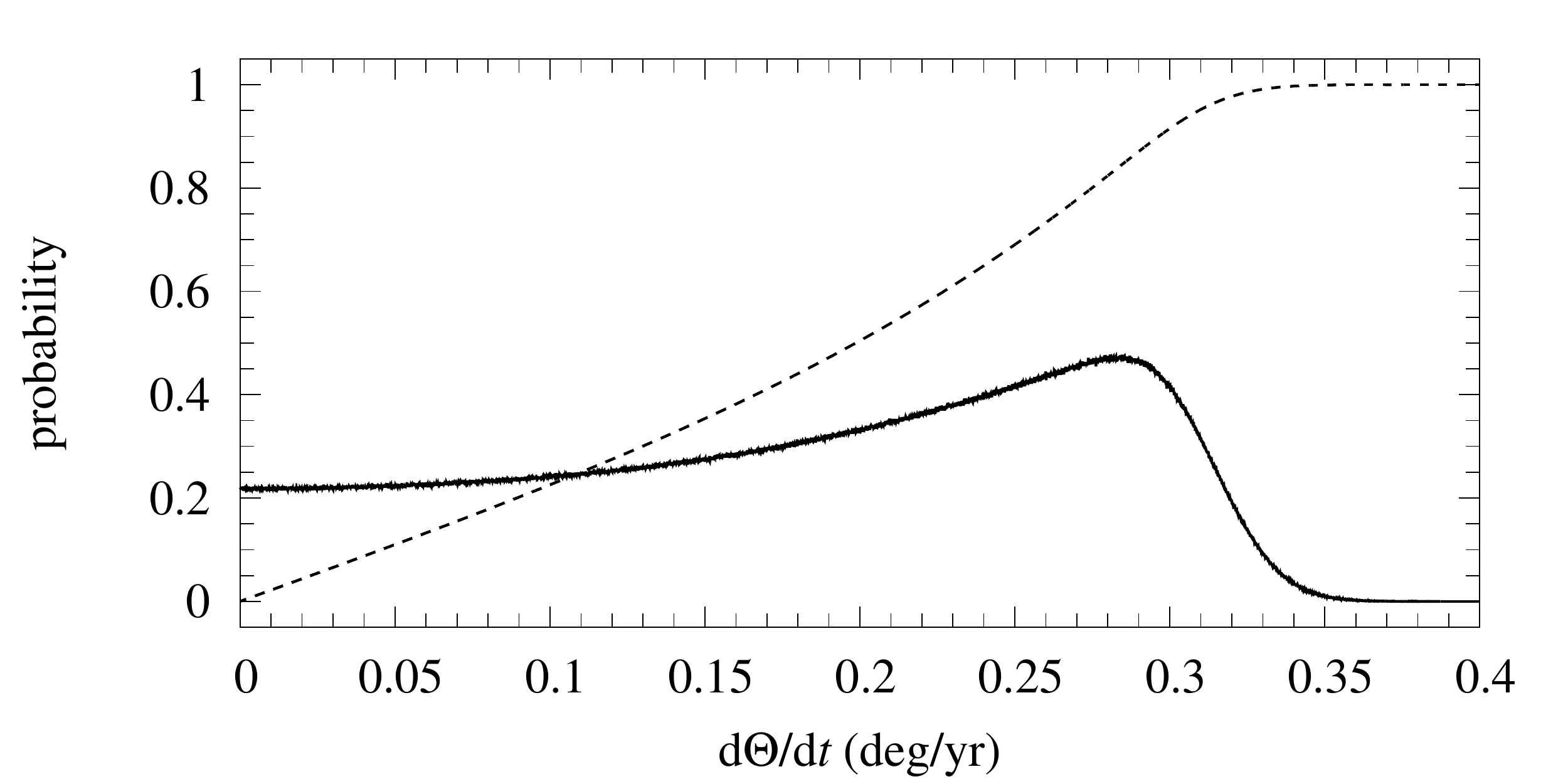, width=8.4cm, clip=}
\caption{As Fig.\,\ref{Prob_a} but for $\dot\Theta$ and circular orbits. 
The PDF was scaled by a factor of 0.1.}
\label{Prob_c}
\end{figure} 

As we did for the circular orbit case, we choose $i$ from an isotropic distribution, and choose specific values for both 
$s$ and $\Delta\gamma$ based on their measured values and assumed Gaussian distributed uncertainties.  We choose the 
longitude of periastron, $\omega$ from a uniform distribution, as is quite reasonable.  

 Finally, we need to choose a
representative value of the outer eccentricity to close the equations.  For this, we utilized the distribution of eccentricities of 
the outer orbits of 222 hierarchical triple systems found in the {\it Kepler} field (B15). This distribution has a maximum 
at about $e=0.35$.  Although the outer orbit of \kic\ is two orders of magnitude larger than is typical for the {\em Kepler} triples, 
we consider the derived distribution as a plausible proxy for quadruple systems consisting of two widely separated close binaries.
Therefore, we chose $e=0.35$ as a statistically representative value for the eccentricity in KIC 7177553.
Equations (6) and (7) are then solved numerically for $a$ by eliminating $E$.

Figures\,\ref{Prob_a} to \ref{Prob_c} show the probability density functions (PDFs) resulting from $10^8$ Monte Carlo trials,
together with the corresponding cumulative probability distributions. 
We obtain highly asymmetric PDFs with maxima at $P=1180$\,yr, $\dot s =1.38$\,mas\,yr$^{-1}$, and
$\dot\Theta=0.28^\circ$yr$^{-1}$ for circular orbits. Allowing for non-circular orbits, the peak in the PDF is
shifted to $P=2060$\,yr and the tail of the PDF extends to more than 10\,000\,yr. 
Table\,\ref{Prob} lists the confidence intervals obtained from the cumulative probability distributions. Here, 
$1\,\sigma$ is a formal designation
corresponding to 67.27\% probability as in the case of a normal distribution.

We also know from the eclipsing binary light curve solution of the 18-day binary, that
its orbital inclination with respect to the observer's line of sight is $i_1=87\fdg 68\pm0\fdg 06$.
If the constituent stars in the two binaries are very similar, as their 
spectra suggest, then the orbital inclination angle of the 16.5\,d binary must be close to $60^\circ$ in 
order for the mass function to yield masses close to 1\,M$_\odot$.  This implies that the two orbital planes 
must be  tilted with respect to each other by at least $\sim$25-30$^\circ$.  

\begin{table}\centering
\caption{$1\,\sigma$ confidence intervals.}
\begin{tabular}{ll}
\hline\hline
$P$          &  1660\,--\,2640 yr \\ 
$\dot s$     & ~0.67\,--\,1.21 mas\,yr$^{-1}$ \\
$\dot\Theta$ & ~0.14\,--\,0.25 $^\circ$\,yr$^{-1}$\\
$P_e$        &  2070\,--\,3550 yr \\
\hline
\end{tabular}\label{Prob}
\tablecomments{Derived from the cumulative probability distributions for circular orbits ($P$, $\dot s$, $\dot\Theta$)
and for an orbit with $e=0.35$ ($P_e$).}
\end{table}

\section{Conclusions}\label{Conc}

The analysis of the radial velocity curves derived from the LSD profiles revealed the SB4 nature of the \kic\ system 
and allowed us to calculate precise orbital solutions for 
the two underlying binaries. The orbital parameters derived from the LSD profiles are consistent with those delivered by the KOREL program. 
We find two eccentric systems having only slightly different  binary periods and consisting of components of almost the same 
masses. The systemic velocities of both systems differ by only 1.5\,\kms.

The analysis of the decomposed spectra showed that all four stars are of comparable spectral type. The errors in the
derived atmospheric parameters are relatively large, however. The reason is that we had to solve for the flux ratios
between the stars as well and thus the number of the degrees of freedom in the combined analysis is high. 
In the result, all atmospheric parameters agree within 1\,$\sigma$ of the error bars. 
What we can say is that all components of the two systems are main sequence 
G-type stars showing abundances close to solar, i.e., we are dealing with four slowly rotating, non-evolved, solar-like stars.
Component $C_1$ has a slightly higher mass and shows the higher continuum flux.
It seems likely that it also has a higher temperature, i.e., that the obtained difference in \teff\ compared to
the other three stars (Table\,\ref{Analysis}) is significant.
The eclipsing binary light-curve analysis of the {\it Kepler} long-cadence photometry confirmed the
spectroscopic results. 

\kic\ most likely belongs to the rare known SB4 quadruple systems consisting of two gravitationally bound binaries.
This assumption is strongly supported by the similar spectral types and apparent magnitudes of the two binaries, as well
as by their small angular separation and similar $\gamma$ velocities.
In particular, several AO surveys \citep[see e.g.][]{2015ApJS..216....7B,2015ApJ...800..138N} show that for nearby objects 
(up to few hundred parsec) almost every pair of stars that is this bright and this closely separated has been shown to be
physically associated.
But as is the case of \object{KIC 4247791} \citep{2012A&A...541A.105L}, the
time span of actual spectroscopic observations is too short to search for any gravitational interaction between the 
two binaries and we neglected such effects in the calculation of  their individual binary orbits. 
 
 The analysis of the O-C values from fitting the RV curves showed that there are no significant changes in
the systemic velocities during the time span of the spectroscopic observations. Only when defining outliers based on a 
2\,$\sigma$-clipping did we observe a decrease of the
systemic $\gamma$-velocity of the 18\,d-binary. Though it is not accompanied by a corresponding increase of the same order
of the $\gamma$-velocity of the other binary, it could be related to a third body in the 18\,d-binary such as the giant planet 
we are searching for.
In a similar way, the analysis of the eclipse timing variations of the eclipsing pair which is described  in B15
 neither proves nor refutes the probable gravitationally bounded quadruple system scenario, but
indicates a possible non-transiting circumbinary planet companion around the 18-day eclipsing pair of \kic.

The outer orbital period of the quadruple system must be longer than 1000 yr and the corresponding
probability distribution 
peaks at about 1200\,yr for circular and at about 2000\,yr when assuming eccentric orbits with $e=0.35$. 
Thus, no orbital RV variations can be measured from spectroscopy in the next decade. 
Further spectroscopic observations are valuable to extend the search for the hypothetical planetary companion, however. 

The proper motion of \kic\ in  RA\,cos(Dec) and Dec is given in the Tycho-2 catalog \citep{2000A&A...355L..27H} as 
6.6 and 14.5\,mas\,yr$^{-1}$, respectively. This is distinctly larger than the expected change of the projected 
separation of the two binaries on the sky per year, $\dot s$, 
and future measurements will show if both binaries have the same proper motion.
But also $\dot s$ as well as a change of the position angle, $\dot\Theta$,
can be measured from speckle interferometry or direct
imaging using adaptive optics in the next few years. The observations with the Keck NIRC2 camera, for example, yield an accuracy
of about 1.5\,mas in radial separation and about 0\fdg2 in position angle which could be sufficient for a secure measurement 
of $\dot s$ and $\dot\Theta$ for an epoch difference of two to three years (see Table\,\ref{Prob} for a comparison).   
Finally, the Gaia satellite mission \citep{2015ASPC..496..121E} with its 24\,$\mu$as astrometric accuracy for $V=15$\,mag stars and 
spatial resolution of 0.1\,mas in scanning direction and 0.3\,mas in cross-direction can resolve the two binaries easily. 
The non-single stars catalog is scheduled to appear in the fourth Gaia release at the end of 2018.

\acknowledgments
This work is based on observations with the 2-m Alfred Jensch Telescope of the
Th\"uringer Landessternwarte Tautenburg. It has made use of data collected by the {\it Kepler} 
satellite mission, which is funded by the NASA Science Mission directorate. 
Some of the data presented herein were obtained at the W.M. Keck Observatory, 
which is operated as a scientific partnership among the California Institute of Technology, 
the University of California and the National Aeronautics and Space Administration. 
The observatory was made possible by the generous financial support of the W.M. Keck Foundation. 
This work has made use of the VALD database, operated at Uppsala University, the Institute of Astronomy RAS 
in Moscow, and the University of Vienna.
The project has been supported by the Hungarian OTKA Grant K113117.

\bibliographystyle{apj}

\newpage ~\newpage ~\newpage
\appendix

\section{Spectroscopic Flux Ratios from the Decomposed Spectra of Multiple Systems}\label{Mathe}

Let $I$ be the observed, composite spectrum showing the lines of $n$ stars, normalized to the total continuum of all $n$ stars,
and $I_k,~k=1..n$, the decomposed spectra of the single stars normalized to the same total continuum,
\begin{equation}
I = \sum_{k=1}^n{I_k}-(n-1)
\end{equation}
or, in line depths with $i = 1-I$,
\begin{equation}
i = \sum_{k=1}^n{i_k}.
\end{equation}
The decomposed spectra shall be fitted by the synthetic spectra $S_k$, which are normalized to the individual continua of the single stars. 
In the ideal case we have $i_k = f_k s_k$, where $s_k = 1-S_K$ are the line depths of the synthetic spectra, $f_k$ is the flux ratio
of star $k$ compared to the total continuum flux, and $\sum{f_k} = 1$. In the following we consider the simplest case 
of constant $f_k$ (the derivation can easily be extended to wavelength-dependent $f_k$ by developing $f_k(\lambda)$
into a polynomial in wavelength $\lambda$). We define
\begin{equation}
\chi^2 = \sqrt{\sum_\lambda{\ds\sum_{k=1}^n{\frac{(i_k-f_ks_k)^2}{\ds\sigma_k^2}}}}
\end{equation}
where $\sigma_k$ is a weighting factor corresponding to the mean uncertainty of the observed spectrum $i_k$.
Setting 
\begin{equation}
f_1 = 1-\sum_{j=2}^n{f_j},
\end{equation}
we minimize
\begin{equation}\label{min}
\sum_\lambda{\left(\frac{\ds 1}{\ds\sigma_1^2}\left[i_1-\left(1-\sum_{j=2}^n{f_j}\right)s_1\right]^2+
\sum_{k=2}^n{\frac{\ds(i_k-f_ks_k)^2}{\ds\sigma_k^2}}\right)}.\vspace{2mm}
\end{equation}
Setting the partial derivatives of Eqn.\,\ref{min} with respect to $f_2$ to $f_n$ to zero yields, with $k = 2..n$,
\begin{equation}
\sum_\lambda{\left(\frac{\ds s_1}{\ds\sigma_1^2}\left[i_1-\left(1-\sum_{j=2}^n{f_j}\right)s_1\right]-
\frac{\ds s_k}{\ds\sigma_k^2}(i_k-f_k s_k)\right)} = 0.
\end{equation}
Sorting by the $f_k$ and dividing by $[s_1^2]/\sigma_1^2$, we obtain, with $k = 2..n$,
\begin{equation}\label{eq}
\frac{\ds\sigma_1^2}{\ds\sigma_k^2}\frac{\ds [s_k^2]}{\ds [s_1^2]}f_k+\sum_{j=2}^n{f_j} = 
1+\frac{\ds\sigma_1^2[s_ki_k]-\sigma_k^2[s_1i_1]}{\ds\sigma_k^2[s_1^2]}
\end{equation}
where the brackets mean the sum over all wavelength bins. Equation\,\ref{eq}
is equivalent to the system of linear equations
\begin{equation}\label{LGS}
\begin{array}{l}
a_2f_2+~~~f_3+....+~~~f_n = h_2\\
~~~f_2+a_3f_3+....+~~~f_n = h_3\\
~~~~~~:~~~~~~~:~~~~~:~~~~~~~:\\
~~~f_2+~~~f_3+....+a_nf_n = h_n\\
\end{array}
\end{equation}
with 
\begin{eqnarray}\label{coeff}
a_k & = & 1+\frac{\ds\sigma_1^2}{\ds\sigma_k^2}\frac{\ds [s_k^2]}{\ds [s_1^2]}\\
h_k & = & 1+\frac{\ds\sigma_1^2[s_ki_k]-\sigma_k^2[s_1i_1]}{\ds\sigma_k^2[s_1^2]}
\end{eqnarray}
The flux ratios follow from solving Eqn.\,\ref{LGS}. Doing this using a grid of atmospheric parameters $p_k$
to calculate different synthetic spectra $s_k(p_k)$ finally yields the optimum set of atmospheric parameters
together with the corresponding optimum flux ratios.  

\section{Measured Radial Velocities}

Table\,\ref{Obs} lists the date of mean exposure, the signal-to-noise ratio of the spectra,  and the RVs 
of the four components of KIC\,7177553 plus their errors in \kms, measured from multi-Gaussian fits to the LSD profiles.

\begin{table*}[h]\centering
\caption{Radial velocities measured for the four components of KIC\,717755.}
\begin{tabular}{crrcrcrcrc}
\hline\hline
BJD & $SN$ & $RV_1$ & $\sigma_1$ & $RV_2$ & $\sigma_2$ & $RV_3$ & $\sigma_3$ & $RV_4$ & $\sigma_4$\\
\hline
2\,457\,170.502322&  82&     3.278&  0.105&  $-$35.959&  0.144&  $-$83.492&  0.123&	57.663&  0.148\\
2\,457\,170.532774& 116&     3.525&  0.101&  $-$36.369&  0.139&  $-$83.282&  0.123&	57.326&  0.144\\
2\,457\,172.437936& 104&    15.358&  0.104&  $-$45.638&  0.129&  $-$45.638&  0.129&	15.358&  0.104\\
2\,457\,172.466640&  99&    15.025&  0.100&  $-$45.516&  0.136&  $-$45.516&  0.136&	15.025&  0.100\\
2\,457\,173.433937& 111&    16.382&  0.139&  $-$50.365&  0.200&  $-$26.624&  0.149&   $-$1.158&  0.177\\
2\,457\,174.380735&  81&    14.755&  0.196&  $-$50.641&  0.374&  $-$14.586&  0.172&  $-$14.586&  0.172\\
2\,457\,174.432658& 116&    16.837&  0.177&  $-$50.461&  0.258&  $-$14.332&  0.116&  $-$14.332&  0.116\\
2\,457\,176.399426&  80&    13.496&  0.159&  $-$47.224&  0.208&      0.959&  0.203&  $-$29.714&  0.188\\
2\,457\,180.403582& 101& $-$15.686&  0.099&  $-$15.686&  0.099&     15.734&  0.175&  $-$45.171&  0.196\\
2\,457\,181.394521&  73& $-$34.550&  0.269&	 5.267&  0.325&     16.495&  0.320&  $-$45.715&  0.320\\
2\,457\,181.423469& 105& $-$35.344&  0.264&	 5.359&  0.304&     16.516&  0.268&  $-$45.907&  0.286\\
2\,457\,184.475914& 133& $-$86.269&  0.121&	58.888&  0.177&   $-$7.206&  0.167&  $-$21.470&  0.143\\
2\,457\,184.505151& 113& $-$85.764&  0.121&	58.062&  0.177&   $-$7.841&  0.175&  $-$20.992&  0.152\\
2\,457\,190.536552& 125&    14.216&  0.129&  $-$48.633&  0.182&  $-$20.204&  0.154&   $-$7.757&  0.188\\
2\,457\,204.439583&  77& $-$28.640&  0.132&   $-$2.398&  0.182&  $-$67.966&  0.166&	41.761&  0.206\\
2\,457\,205.466712& 115&  $-$8.225&  0.175&  $-$23.512&  0.219&  $-$45.066&  0.181&	17.131&  0.213\\
2\,457\,206.404903& 148&     2.022&  0.087&  $-$31.511&  0.127&  $-$31.511&  0.127&	 2.022&  0.087\\
2\,457\,206.508631& 141&     1.481&  0.089&  $-$36.269&  0.311&  $-$26.209&  0.273&	 1.481&  0.089\\
2\,457\,207.415514& 106&     9.428&  0.138&  $-$43.305&  0.205&  $-$14.533&  0.114&  $-$14.533&  0.114\\
2\,457\,207.519392& 111&     9.370&  0.144&  $-$44.657&  0.213&  $-$14.633&  0.106&  $-$14.633&  0.106\\
2\,457\,207.547575&  92&     8.936&  0.157&  $-$44.974&  0.237&  $-$14.606&  0.116&  $-$14.606&  0.116\\
2\,457\,210.467746& 115&    16.897&  0.228&  $-$51.306&  0.227&      5.854&  0.258&  $-$35.446&  0.227\\
2\,457\,210.496392&  99&    16.690&  0.224&  $-$51.347&  0.230&      5.724&  0.261&  $-$35.509&  0.223\\
2\,457\,234.396006&  87& $-$15.123&  0.052&  $-$15.123&  0.052&  $-$15.123&  0.052&  $-$15.123&  0.052\\
2\,457\,235.417500&  87& $-$36.495&  0.289&	 4.618&  0.239&  $-$51.053&  0.289&	23.735&  0.236\\
2\,457\,236.407512& 133& $-$61.497&  0.098&	31.786&  0.117&  $-$83.062&  0.086&	57.428&  0.113\\
2\,457\,237.460821& 138& $-$88.241&  0.105&	59.267&  0.168&  $-$70.586&  0.117&	43.846&  0.173\\
2\,457\,238.422082& 105& $-$87.637&  0.210&	59.069&  0.290&  $-$48.249&  0.255&	21.223&  0.305\\
2\,457\,240.529068&  80& $-$26.729&  0.965&   $-$1.422&  0.402&  $-$13.932&  0.295&  $-$13.932&  0.295\\
\hline
\end{tabular}\label{Obs}
\end{table*}

\end{document}